\documentclass{article}

\PassOptionsToPackage{square,numbers,sort,compress}{natbib}
\usepackage{natbib}
\usepackage[preprint]{neurips_2024}

\usepackage[utf8]{inputenc} 
\usepackage[T1]{fontenc}    
\usepackage[colorlinks=true, urlcolor=blue, citecolor=blue, linkcolor=blue]{hyperref}      
\usepackage{url}            
\usepackage{booktabs}       
\usepackage{amsfonts}       
\usepackage{nicefrac}       
\usepackage{microtype}      
\usepackage{xcolor}         
\usepackage{doi}
\usepackage{graphicx}
\usepackage{graphics}
\usepackage{caption}
\usepackage{subcaption}
\usepackage{multicol}
\usepackage{colortbl}
\usepackage{longtable}
\usepackage{xurl} 
\usepackage{hyperref}
\usepackage{tabularx}
\usepackage{soul}
\usepackage{tcolorbox}
\usepackage{lipsum}  
\usepackage{scalerel,xparse}

\title{\textsc{Standardizing Intelligence:}\\Aligning Generative AI for Regulatory and Operational Compliance}

\author{%
  Joseph Marvin Imperial$^{1,2}$~~~~~Matthew D. Jones$^{3}$~~~~~Harish Tayyar Madabushi$^{1,2}$
  \\
  \\
  $^{1}$UKRI CDT in Accountable, Responsible and Transparent AI\\
  $^{2}$Department of Computer Science~~~~$^{3}$Department of Life Sciences\\
  University of Bath, UK\\
  \\
  \texttt{jmri20@bath.ac.uk}~~~~\texttt{prpmdj@bath.ac.uk}~~~~\texttt{htm43@bath.ac.uk}
}

\begin{document}

\maketitle


\vspace{-2em} 
\begin{center}
\begin{minipage}[c]{0.4\textwidth}
\centering
\includegraphics[width=0.8\textwidth]{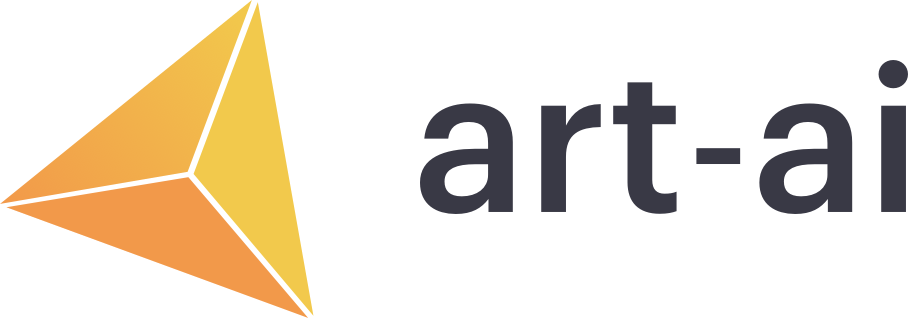}
\end{minipage}%
\hspace{-2em}
\begin{minipage}[c]{0.4\textwidth}
\centering
\includegraphics[width=0.8\textwidth]{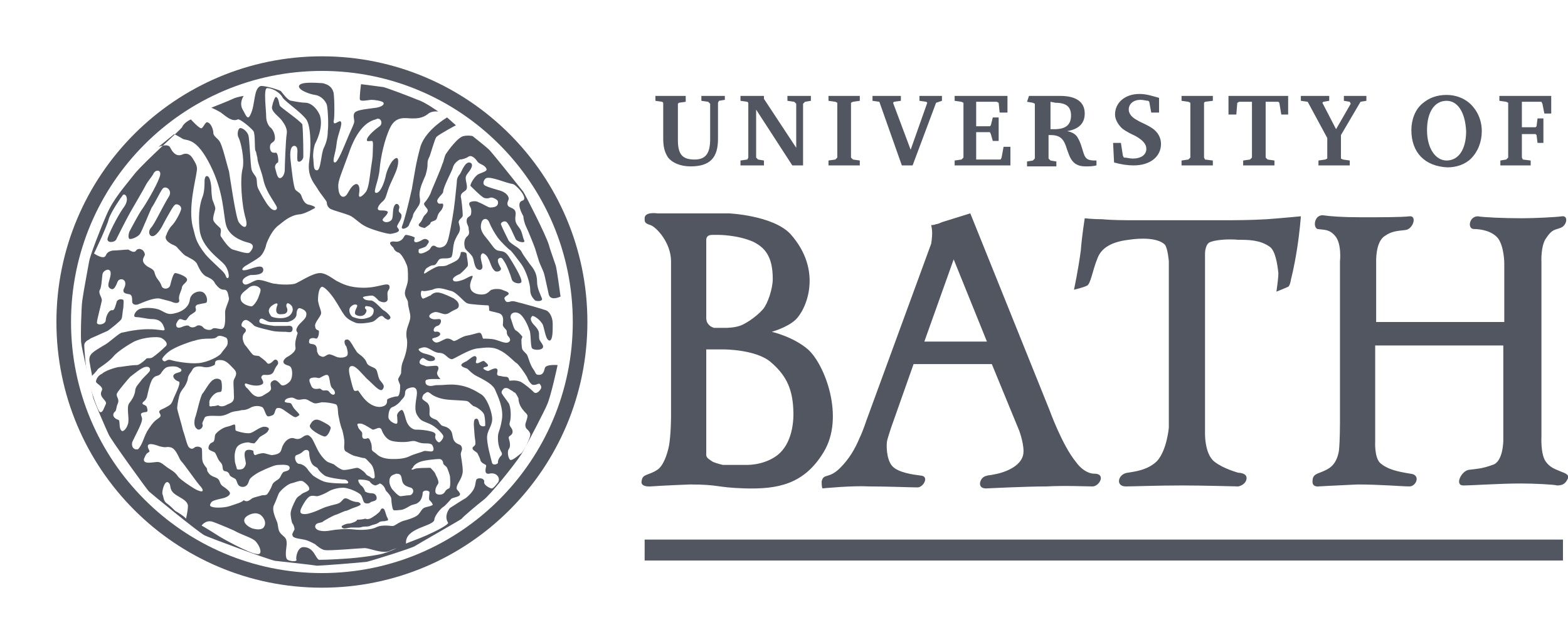} 
\end{minipage}
\end{center}
\vspace{1em} 

\definecolor{baseline}{RGB}{222, 246, 214} 
\definecolor{specialized}{RGB}{199, 239, 181} 
\definecolor{advanced}{RGB}{165, 215, 142} 
\definecolor{adaptive}{RGB}{104, 196, 109} 

\definecolor{minimal}{RGB}{255, 224, 224} 
\definecolor{moderate}{RGB}{244, 204, 204} 
\definecolor{high}{RGB}{239, 167, 167} 
\definecolor{extreme}{RGB}{217, 117, 117} 

\newcommand{\baseline}{\colorbox{baseline}{Baseline}}
\newcommand{\specialized}{\colorbox{specialized}{Specialized}}
\newcommand{\advanced}{\colorbox{advanced}{Advanced}}
\newcommand{\adaptive}{\colorbox{adaptive}{Adaptive}}
\newcommand{\minimal}{\colorbox{minimal}{Minimal}}
\newcommand{\moderate}{\colorbox{moderate}{Moderate}}
\newcommand{\high}{\colorbox{high}{High}}
\newcommand{\extreme}{\colorbox{extreme}{Extreme}}

\begin{abstract}
Technical standards, or simply \textit{standards}, are established documented guidelines and rules that facilitate the interoperability, quality, and accuracy of systems and processes. In recent years, we have witnessed an emerging \textit{paradigm shift} where the adoption of generative AI (GenAI) models has increased tremendously, spreading implementation interests across standard-driven industries, including engineering, legal, healthcare, and education. In this paper, we assess the \textit{criticality levels} of different standards across domains and sectors and complement them by grading the current \textit{compliance capabilities} of state-of-the-art GenAI models. To support the discussion, we outline possible challenges and opportunities with integrating GenAI for standard compliance tasks while also providing actionable recommendations for entities involved with developing and using standards. Overall, we argue that \textit{aligning GenAI with standards through computational methods can help strengthen regulatory and operational compliance}. We anticipate this area of research will play a central role in the management, oversight, and trustworthiness of larger, more powerful GenAI-based systems in the near future.
\end{abstract}

\section{Introduction}
\label{sec:intro}


\begin{figure}[!t]
    \centering
    \includegraphics[width=0.70\linewidth, trim=0 7 0 0, clip]{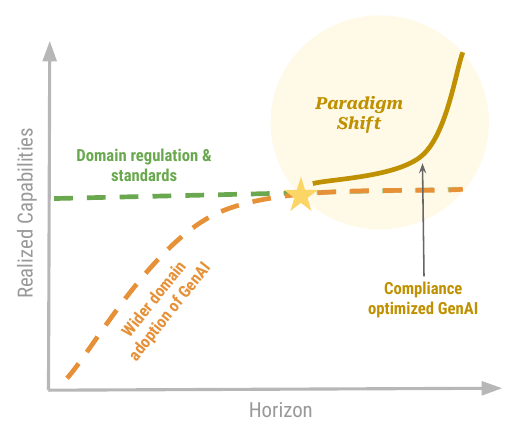}
    \caption{We describe an emerging \textbf{paradigm shift} where domain experts from interdisciplinary areas such as education, engineering, and healthcare are using advanced generative AI models (e.g., GPT-4) to assist them with regulatory and operational compliance through standards. We pattern the temporal observation of the paradigm shift within the near-to-midterm realized capabilities of GenAI as described in \citet{pmlr-v235-eiras24b}.}
    \label{fig:paradigm-shift}
\end{figure}


The industries of the modern world rely on systematic processes for the efficient production of goods and delivering services guided through \textbf{standards}. According to the International Organization for Standardization (ISO)\footnote{\url{https://www.iso.org/standards.html}}, standards refer to a general form of documented specifications, rules, and norms specialized across various domains and sectors such as healthcare, education, engineering, science, communication, security and defense. 
For example, in the aerospace engineering domain, any technical instruction manual produced must conform to recognized technical language standards such as ASD-STE100 Simplified Technical English (STE)\footnote{\url{https://www.asd-ste100.org/}} developed and maintained by the Aerospace, Security and Defence Industry Association of Europe (ASD, formerly AECMA). In the education and language proficiency assessment domain, on the other hand, teachers in charge of developing curriculum materials must follow education standard frameworks such as the Common European Framework of Reference for Languages (CEFR)\footnote{\url{https://www.coe.int/en/web/common-european-framework-reference-languages}} in Europe or the Common Core State Standards (CCS)\footnote{\url{https://www.thecorestandards.org/read-the-standards/}} in North America to produce high-quality classroom content.

Standards are typically composed of carefully defined technical \textit{specifications} for measurements, design, and performance quotas that can be used to check or validate regulatory and operational compliance while preserving interoperability, quality, and accuracy \cite{astm2025}. In recent years, there has been observable interest from users across various domains in adopting these instruction-following generative AI (GenAI) technologies, such as ChatGPT, due to their documented capabilities, including the ability to follow complex instructions and generate human-like writing. For example, a survey by the Department of Education revealed that 62\% of primary and secondary teachers in the UK have reported using GenAI tools to create new educational content and lessons for classroom use\footnote{n = 230. Data gathering was conducted around August and September 2023 with respondents from 23 educational settings.} \cite{DfE2024GenAI}. Complementary to this, recent empirical works on benchmarking GenAI models for automatic content generation using CEFR and CCS standards as references for control show that off-the-shelf commercial and open-weight LLMs such as Llama2 \cite{touvron2023llama} and GPT-4 \cite{achiam2023gpt} can be systematically steered to produce high-quality content through methods such as in-context learning (ICL) and reinforcement learning (RL) while preserving high automatic and human expert evaluations \cite{imperial-etal-2024-standardize, malik-etal-2024-tarzan}. Thus, this promising research direction of aligning GenAI models with standards underscores the need for greater attention from both the AI and interdisciplinary research communities to examine how GenAI is transforming regulatory and operational compliance across standard-driven domains.

In this paper, we analyze the changing landscape---an emerging \textit{paradigm shift}---of regulatory and operational compliance through standards across various sectors and domains. We propose a joint \textbf{\textsc{Criticality and Compliance Capabilities Framework (C3F)}} for assessing the measured capabilities of 15 recent and community-recognized foundational and specialized GenAI models for standard-compliance tasks and the criticality levels of 34 standards from various domains based on their sensitivity and potential consequences in case of non-compliance. We cover a variety of case studies supporting the paradigm shift, including healthcare, education, safety, finance, and engineering. We outline possible challenges and opportunities with learning standard compliance using GenAI models, the benefits if done successfully, and recommendations for various stakeholders involved. Finally, we take the following position that \textbf{aligning GenAI with standards through computational methods can help strengthen regulatory and operational compliance}. Thus, leading to enhanced control, oversight, and trust among these systems in real-world settings.

\section{The Regulatory and Operational Compliance Landscape}
\label{sec:background}

To set the stage, we discuss preliminary information on the common definitions and processes comprising the development of standards, including common characteristics and entities involved in their conception and development.

\subsection{Defining A Standard}
Standards are established to provide specifications for measuring quality and proof of compliance with regulations. In line with this, we consider standards as \textit{expert-defined} documents since one typically needs substantial knowledge within a domain or sector to propose measurable requirements and regulatory orders for compliance from users, industries, and organizations \cite{demortain2008standardising}. The typology of standards can be generalized into either \textit{product-based} standards that specify target characteristics for physical or digital products or \textit{non-product} standards that govern and specify target efficiency, operation, and performance-based measures for processes and services \cite{nsf2018}.

To cater to the diverse overlapping notions of standards, in this paper, we do not restrict the scope of standards to those created by international and regional standard developing organizations (SDOs) such as ISO, IEEE, ETSI, CEN-CENELEC, or NIST. We also include formally documented rules coined under related terms like \textit{frameworks}, \textit{guidelines}, and \textit{checklists}, which are often used in interdisciplinary domains as forms of standards themselves since they observe objectively similar nature and usage. In summary, to unify the overlapping characteristics, a documented set of guidelines can be considered a standard if it conforms to the following below:

\textbf{Anatomy}: Standards are composed of well-defined specifications and procedures documenting measurable requirements for a given product or process.

\textbf{Purpose}: Standards serve specific purposes for various domains and sectors, such as introducing a formal language of communication, defining recognized procedures, ensuring compatibility checks, specifying performance requirements, and assuring compliance with regulations.

\textbf{Recognition}: Standards are developed and recognized by members and constituents of a private or public domain, sector, organization, or regulatory body.


\subsection{Standards as Co-Regulation and Co-Integration Tools}
A standard may be developed publicly or privately through initiatives by the government or regulatory bodies, unions, organizations, and expert groups. In legislation, a standard may be paired with specific laws as a form of a \textbf{co-regulation tool} which, if successfully fulfilled, can serve as a \textit{form of compliance with related state or nation-wide jurisdictions} \cite{pouget2024ai}. In this case, a regulatory body may appoint one or more SDOs to initiate the workflow of creating a specific standard that contains the legal requirements that must be included in line with the law. An example of these standards includes the well-known ISO/IEC 27001:2022 which defines measures for Personally Identifiable Information (PII) controllers and processors of any information security management system (ISMS) in response to the EU General Data Protection Regulation (GDPR) for data privacy and protection \cite{iso27001_2022}.

On the other hand, standards developed through non-legislative initiatives can serve as a \textbf{co-integration tool} which focuses on organization- or community-wide \textit{interoperability and harmonization of systems and processes}. A well-known example is the Web Standards developed by the World Wide Web Consortium (W3C)\footnote{\url{https://www.w3.org/standards/}} which maintains all technical specifications, guidelines, and protocols for web-based technologies including HTML, CSS, and XML. These standards and technologies are globally recognized and form the core building blocks of the Web or Internet. 


\section{Paradigm Shift with GenAI}
\label{sec:paradigm_shift}

A \textbf{paradigm shift} occurs when a dominant standard practice becomes incompatible due to some emerging technological phenomena, facilitating the adoption of new forms of conceptualization, practices, or paradigms \cite{kuhn1970nature, tapscott1994paradigm}. The current state-of-the-art GenAI models are known to exhibit remarkable capabilities across a wide range of generative tasks. In particular, one of the most useful and powerful skills a model can learn is the ability to \textit{follow complex human instructions} from prompts \cite{weifinetuned2022,ouyang2022training,chung2024scaling}. Standards are composed of technical specifications which, at their core, can also be considered a set of instructions. As such, it was not long until users and practitioners knowledgeable of standards in their specific domains and sectors started exploring and reframing these specifications as instructional prompts for GenAI models (e.g., GPT-4) to assist with compliance-based tasks. We consider this phenomenon as an \textit{{emerging paradigm shift}} in standards and regulatory compliance, as shown in Figure~\ref{fig:paradigm-shift}. This paradigm shift is introduced by the rise of instruction and preference-optimized GenAI models that can follow specifications derived from standards through well-structured prompting techniques and domain-specific fine-tuning.


In this section, we further discuss specific cases from the literature in relation to the paradigm shift observed in two major aspects: 1) \textbf{conformity assessment} practices with standards and 2) \textbf{generating standard-aligned content} across various domains and sectors.

\subsection{Shift in Standard Conformity Assessment}
Conformity assessment, in relation to standards, pertains to how implementing organizations and users measure the level to which their products or services meet the requirements of the standard itself. This process is often formally known as \textit{certification} and is extremely variable and dependent on several factors, including the level of conformity required by the standard and common assessment norms in specific domains or sectors  \cite{iso_conformity_assessment}. For some standards-driven sectors such as pharmaceuticals, automotive, and energy industries, certifications are a legal or contractual requirement. For non-regulatory and operational standards, certifications are less common, and conformity can often be assessed through various means, including using third-party evaluator software, hiring trained expert evaluators, or self-assessment by learning the standards from publications or documentation releases. We highlight notable works across various domains in automating conformity assessment with GenAI below:

\noindent \textbf{Data Privacy Laws}. Well-known data privacy laws such as the General Data Protection Regulation (GDPR)\footnote{\url{https://gdpr-info.eu/}} \cite{regulation2016regulation} for the EU and the Health Insurance Portability and Accountability Act (HIPAA)\footnote{\url{https://www.hhs.gov/hipaa/index.html}} \cite{act1996health} for the US have served as common ground for optimizing GenAI models in terms of compliance checking due to the availability of data. \citet{fan-etal-2024-goldcoin} proposed the GoldCoin framework, which leverages contextual integrity theory to build synthetic case scenarios showing compliance and violations of the HIPAA Privacy Rule. The works of \citet{zoubi-etal-2024-privat5} and \citet{zhu2024legilm} introduced PrivT5 and LegiLM, new specialized finetuned models trained from compilations of GDPR-related legal content such as case laws and data-sharing contracts and reported state-of-the-art performance in legal compliance tasks in NLP. 

\noindent \textbf{Financial and Accounting Report Standards}. 
The International Financial Reporting Standards (IFRS) provides a standardized method of evaluating a company's financial performance for compliance across national and international regulations \cite{posner2010sequence}. Auditing financial documents for compliance is considered labor-intensive, and the exploration of AI-driven solutions has been evident in recent years \cite{albuquerque2024exploring}. The work of \citet{berger2023towards} in collaboration with PwC Germany reported the effectiveness of GPT-4 for compliance validation of text sections from financial reports concerning IFRS and Germany's Handelsgesetzbuch or Commercial Code (HGB) through template-based prompting while noting the need for a major improvement in robustness before deploying to real-world scenarios.

\noindent \textbf{Operational Design Standards for Driving Autonomous Systems}. Beyond text-based applications, GenAI models have also been explored and have shown promising results for compliance assessment in multimodal settings. The work of \citet{hildebrandt2024odd} examined the use of OpenAI's ChatGPT-4V \cite{openai_gpt4v_system_card} and Vicuna \cite{chiang2023vicuna} integrated in a pipeline called ODD-diLLMma to check the compliance of compiled sensor image data from self-driving cars with respect to Operational Design Domains (ODDs). ODDs are documented standards provided by manufacturers (e.g., Tesla or GMC) describing specific conditions under which a self-driving car may operate safely and within its designed function (e.g., \textit{vehicle must not be driven at night}). The proposed \textsc{ODD-diLLMma} pipeline is considered the first to automate the compliance checking of ODDs using multimodal LLMs with high accuracy between \~85\%-94\% across 11 weather, environment, and roadway characteristic dimensions.

\subsection{Shift in Standard-Aligned Content Generation}
Automatically generating text- or image-based content that adheres to a specific set of detailed specifications is considered a challenging task, even for AI-based models. State-of-the-art language models like GPT-4 can be prompted to create content using seed topics in which specific syntactic and semantic characteristics can be directed 
\cite{pu-demberg-2023-chatgpt,zhou2023controlled}. Vision-based and multimodal models have also demonstrated the same level of controllability through prompting, particularly in models like DALL-E \cite{betker2023improving} and Stable Diffusion \cite{rombach2022high}. This degree of controllability via simple interactions through a chat interface, which can easily be utilized by users, has been pivotal to the applications of these models across various domains and sectors. We emphasize previous studies that focused on improving GenAI models' capabilities to automatically generate content that conforms to the standards below:

\noindent \textbf{Education and Language Proficiency Frameworks}.
Content-based standards serve as a meter to ensure that classroom resources, such as reading and activity books, meet certain research-based quality criteria \cite{la2000state,sadler2017academic}. An example of a content standard is the Common European Framework of Reference for Languages (CEFR), which is one of the most used resources for automatic educational content generation tasks using LLMs. The combined recent works of \citet{imperial-etal-2024-standardize}, \citet{malik-etal-2024-tarzan}, and  \citet{glandorf-meurers-2024-towards} have all explored a wide range of LLMs, including Llama, GPT-4, and Mistral, using specifications from CEFR in prompts to steer for desired granularities, including target complexity, grammar rules and structure, and levels of meaning. Experts in language testing using CEFR in \citet{imperial-etal-2024-standardize} have also given positive feedback on how GPT-4 can achieve a certain level of completeness, fluency, and coherence in generated texts.

\noindent \textbf{Medical Reporting and Appraisal Standards}.
High-quality documentation and appraisal in the medical literature are driven by checklists and reporting standards. \citet{sanmarchi2024step} explored ChatGPT's capabilities to reformulate the STROBE checklist \cite{von2007strengthening} to analyze epidemiology studies related to COVID-19 vaccinations in 68 countries. The results of the study support ChatGPT's potential as an assistant in setting up epidemiological observational research but caution against its tendency to produce inconsistent responses when analyzing methods. In the same vein, \citet{muluk2024enhancing} also used ChatGPT for customizing checklists related to managing patient-specific musculoskeletal injections that also conform to the METRICS standard \cite{sallam2024preliminary}. The study echoed the considerable potential of GenAI models like ChatGPT for streamlining easily verifiable aspects in clinical practices, such as preparing medical checklists, but emphasized the importance of expert oversight.

\noindent \textbf{Industry Safety Policies}.
Recent works have explored integrating industry safety policies into GenAI models to improve their capability to generate content that adheres to these specifications. An example of this is the Deliberative Alignment training paradigm used in OpenAI's \textsc{o}-series model \cite{guan2024deliberative}. In this method, LLMs are optimized to interpret the company's safety specifications and reason over them when responding to potentially harmful prompts. Another advantage of this method is that the models have been optimized to identify which policy specifications might be relevant to produce a compliant response, rather than going through the full copy of the policy at every iteration. Likewise, the work of \citet{zhang2024controllable} also observes a similar approach to safety policy alignment but focuses on controlling the levels of safety by retraining models across different providers (e.g., \textit{safety policies for generating realistic dialogues for video games can be relaxed to allow cursing}).


\section{\textsc{Criticality and Compliance Capabilities Framework (\textsc{C3F})}}
\label{sec:framework}

\begin{figure}[!t]
    \centering
    \includegraphics[width=0.99\linewidth]{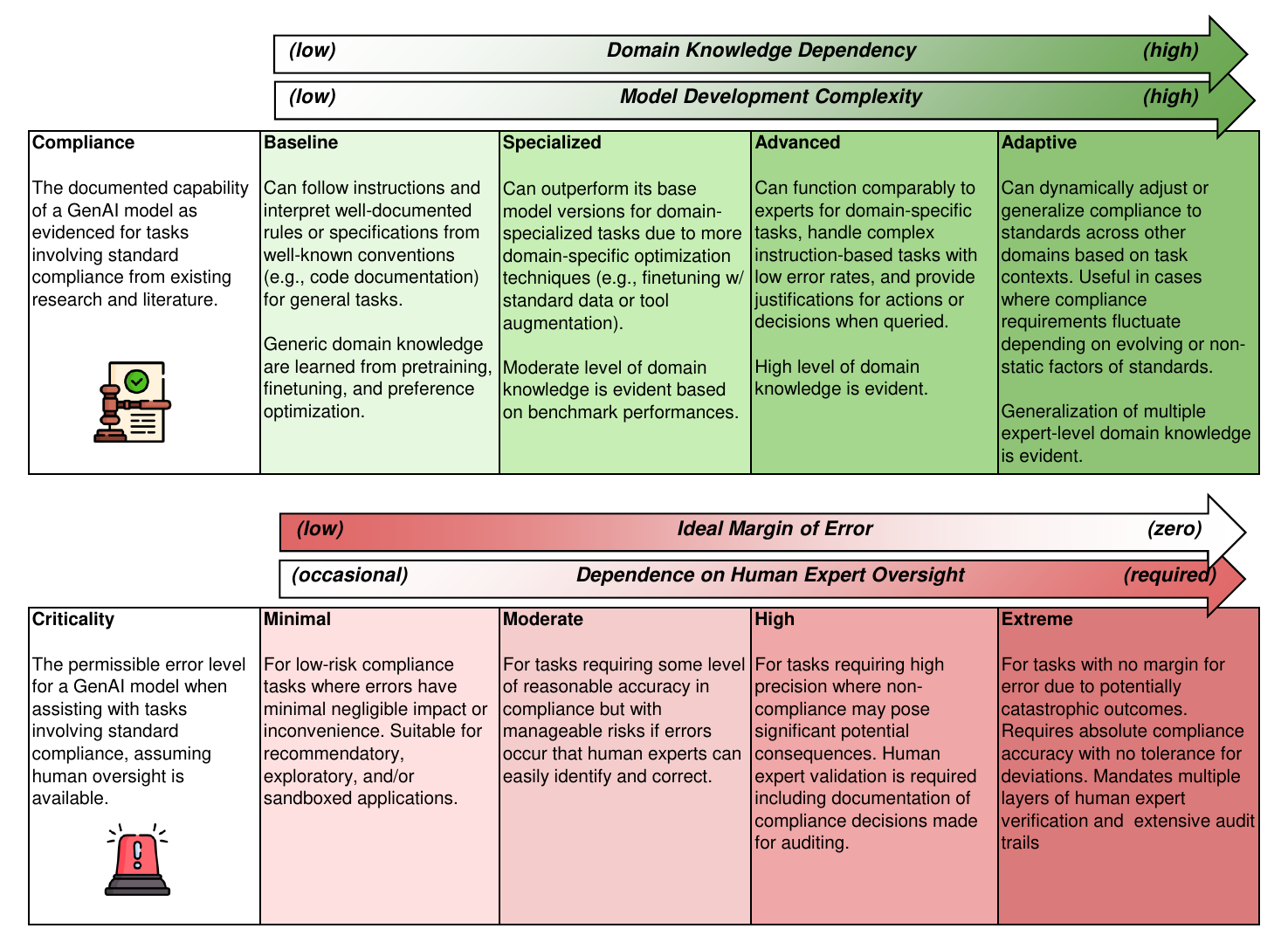}
    \caption{\textbf{The \textsc{Criticality and Compliance Capabilities Framework (C3F)}.} We introduce a joint framework for assessing the current state-of-the-art foundational and specialized text and image-based GenAI models based on their (\textbf{Top}) documented \textit{compliance capabilities} for generating content that aligns with standards, as well as (\textbf{Bottom}) the estimated \textit{criticality} of standards from various domains and sectors based on the permissible error level a GenAI model can commit and the potential consequences in the case of non-compliance.}
    \label{fig:framework}
\end{figure}

The level of sensitivity and compliance requirements vary by standard and depend on its purpose and scope (see discussion on compliance in Section~\ref{sec:paradigm_shift}). For example, coding standards such as Python Enhancement Proposals (PEP) are less sensitive and critical than healthcare standards like the HIPAA Privacy Rule. Thus, for researchers exploring areas using GenAI as an assistant with compliance-based tasks, it is important to know \textit{both} the documented capabilities of the GenAI models they plan to use for experiments \textit{and} the target level of standard compliance they use as a benchmark for success or failure. To bridge this gap, we propose \textsc{C3F}, a joint \textsc{Criticality and Compliance Capabilities Framework} in Figure~\ref{fig:framework} to classify the current capabilities of modern GenAI models to follow compliance with standards as well as assess the different levels of criticality of standards across domains and sectors. We provide a more in-depth discussion of the two components of \textsc{C3F} in the succeeding sections.

\subsection{Classification of GenAI by Documented Compliance Capabilities}

We define {\textbf{compliance capabilities}} as an aggregation of a GenAI model's documented capabilities for compliance-based tasks across various publications and recognition from the interdisciplinary community. For \textsc{C3F}, we propose a four-level assessment scheme to measure a GenAI model's compliance capabilities in an increasing linear fashion as seen in Figure~\ref{fig:framework}. 

\noindent \textbf{Compliance Capabilities Grading Scheme}. For the \baseline level, we consider GenAI models that have been instruction-tuned as a \textit{minimum qualification skill} for assessing compliance capabilities since the core nature of standards is to conform to their specifications. Instruction-tuned GenAI models (particularly LLMs) can pick up generic domain knowledge from the massive datasets often used for pretraining and additional output optimizations through instruction tuning and preference optimization. This is particularly evident from tasks such as code generation \cite{siddiq2023lightweight,beer2024examination} and health-related checklists \cite{sanmarchi2024step,muluk2024enhancing}. Succeeding levels, including Specialized and Advanced, require further evidence of domain knowledge expertise that outperforms Baseline-level models. GenAI models classified as \specialized are typically those that have been additionally fine-tuned with domain-specific gold-standard datasets such as clinical guidelines \cite{chen2023meditron} and examples of regulatory compliant documents \cite{fan-etal-2024-goldcoin}. On the other hand, \advanced GenAI models are those that can be considered equal to domain experts for the task of standard compliance while also being capable of justifying or reasoning over decisions with constraints from standards. Lastly, we consider \adaptive as the final level a GenAI model can obtain, which should exhibit the highest form of capability via generalization of multiple expert-level knowledge across domains. No existing GenAI model is currently classified under this level.

\noindent \textbf{Assessment}. We used the compliance capabilities component of \textsc{C3F} to assess 15 foundational and specialized GenAI models both for text and image as shown in Table~\ref{tab:model_classification} in Appendix~\ref{app:framework_tables} which includes information on their respective domains (in the case of Specialized models) and accessibility. We note that only the o-series models (o1 and o3) from OpenAI have been documented to exhibit the required capabilities to be classified under the Advanced category, as evidenced by their Deliberative Alignment study, which shows how LLMs can be trained to reason and select which applicable specifications from safety policies should be used to generate a safe response \cite{guan2024deliberative}. Recently released models, such as DeepSeek-R1 \cite{deepseek2025}, are currently classified as Baseline and can be updated upon publication of literature documenting if their compliance capabilities can quality for the Advanced level.

\noindent \textbf{Observation}. We highlight two complementing observational points---(1) domain knowledge dependency and (2) model development complexity---in the compliance capabilities of the framework (reflected as two green gradient arrows in Figure~\ref{fig:framework}). The first point, \textit{domain knowledge dependency}, describes a direct relationship between the documented compliance capabilities of GenAI models and their evidenced domain knowledge. This is a straightforward observation as Specialized (and higher level) models will typically outperform their Baseline versions for domain-specific compliance tasks. The second point, \textit{model development complexity}, reflects a similar direct relationship where higher compliance capabilities of GenAI models demands increasing complexity and costly data curation and training procedures. As a consequence, AI-based companies with higher financial resources such as Google, Meta, and OpenAI typically lead the development of more compute-heavy models.

\subsection{Classification of Standards by Criticality Levels}

We define the \textbf{criticality levels} of standards as a measure of their sensitivity, which can be determined by the allowable margin of error for a hypothetical GenAI model assisting with standard compliance tasks. For \textsc{C3F}, in parallel with compliance capabilities, we also propose a similar four-level assessment scheme illustrating the reduction of allowable errors as the criticality of standards increases as shown in Figure~\ref{fig:framework}. 

\noindent \textbf{Criticality Level Grading Scheme}. We consider \minimal criticality level as the least sensitive and can be used for GenAI-based experiments without requiring in-depth domain knowledge or expert oversight. Non-compliance can also be easily detected with existing rule-based software. This includes standards such as coding conventions (e.g., Python Enhancement Proposals) or writing and formatting guidelines (e.g., Plain Language, SMILES in Chemistry). For \moderate, there are potential risks associated with non-compliance but they can easily be managed and corrected by human experts. Examples in this category include most non-regulatory standards, standards with variations across domains, and standards developed by independent, private-sector organizations primarily used for interoperability, such as PRISMA for reporting systematic review papers and IFRS Accounting Standards in finance. Standards classified under  \high are those that require high levels of accuracy and may pose significant consequences in case of non-compliance. Most patient-facing healthcare standards classified in this category include the SPIRIT Checklist, which is used for transparency of clinical trial protocols; GDPR and HIPAA for data protection and privacy; and the USDA Food Safety Documentation. Lastly, the highest criticality level a standard can be classified as is \extreme, which is reserved for situations with zero margin of error allowed, and non-compliance may result in catastrophic and potentially irreversible consequences. This includes standards under the chemical, biological, radiological, and nuclear (CBRN) umbrella, such as the Safety Standards developed by the International Atomic Energy Agency and the Joint Operating Principles for Emergency Services, which pertain to documenting CBRN-related emergency responses. Such a high degree of sensitivity and criticality is necessary to include, as works on GenAI, particularly LLMs, are already gaining research attention and preliminary works \cite{de2024classification,hirata2024generative}.

\noindent \textbf{Assessment}. We used the standard criticality level component of \textsc{C3F} to assess 34 globally recognized standards (including guidelines, checklists, and policies) from a wide range of domains and sectors. While the level of compliance with a certain standard can be deduced by reading its respective documentation and release reports, collecting the opinions of domain experts can justify its classification based on criticality from \textsc{C3F}, which is a normal practice in the conventional standards development process shown in Figure~\ref{fig:standard-hierarchy} in Appendix~\ref{app:framework_tables}. In assessing the standards, we consider two things: the \textit{consequences of harm} and the \textit{scale of harm}. The former describes the level of potential damage that non-compliance with standards can trigger, while the latter considers the number of people who might be harmed by an error caused by non-compliance. For example, an error in a patient-facing scenario in healthcare might endanger one person, but an error in a nuclear energy scenario might endanger or kill thousands. Both can lead potentially to death, but the scale varies between the two. For standards classified under the domains of healthcare and engineering in Table~\ref{tab:standards_classification}, we conversed with two practitioners from our university network who have experience using the standard and obtained their assessments based on \textsc{C3F}.

\noindent \textbf{Observation}. We also highlight two observational points---(1) ideal margin of error and (2) dependence on human expert oversight---in terms of criticality levels proposed in the framework (reflected as two red gradient lines). In this case, however, the two points are opposites. The \textit{ideal margin of error} should decrease from low to zero as criticality levels increase, particularly for standards rated High and Extreme, to avoid the potential consequences of non-compliance. The \textit{dependence on human expert oversight}, on the other hand, is directly proportional and should increase from occasional (for Minimal criticality) to required (for High and Extreme) as criticality levels also increase.

\section{Challenges and Opportunities}
\label{sec:challenges}

\NewDocumentCommand\emojiplus{}{
    \scalerel*{
        \includegraphics{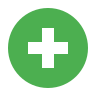}
    }{X}
}

\NewDocumentCommand\emojiinfo{}{
    \scalerel*{
        \includegraphics{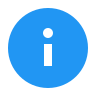}
    }{X}
}

\NewDocumentCommand\emojipeople{}{
    \scalerel*{
        \includegraphics{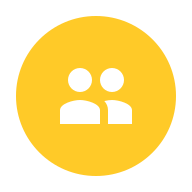}
    }{X}
}

\NewDocumentCommand\emojitool{}{
    \scalerel*{
        \includegraphics{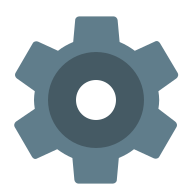}
    }{X}
}

\newcommand{\positiveOpp}[1]{\emojiplus #1}
\newcommand{\challengeOpp}[1]{\emojiinfo #1}
\newcommand{\technicalOpp}[1]{\emojitool #1}
\newcommand{\recommendationOpp}[1]{\emojipeople #1}

To complement the categorization of standards by criticality levels in Section~\ref{sec:framework}, we provide an extensive discussion of the technical aspects that make the process of aligning GenAI models with standards both challenging and promising, and highlight the advantages that standards can introduce to existing GenAI-based systems

\subsection{Complexities of Standards}
\textbf{\challengeOpp{Standards Are Living Documents}} Standards can be considered as \textit{living documents} where their contents can be revised, changed, or updated at any point in time by their developers. SDOs typically perform periodic reviews of their standards (5 in the case of ISO), which are driven by factors such as industry changes, the introduction of new methods or products, and significant technological advances in the field \cite{eetimes_when_standards_change, bsigroup_standards_terminology}. Other forms of standards we consider in this work, such as organizational policies and guidelines, can be updated as needed. As such, architectures for standard-aligned GenAI model pipelines should be designed to adapt dynamically, accommodating both minor and major changes in a standard's specifications for compliance-based tasks. A good example of this is ISO/IEC 22989:2022, where new AI-related terminologies and concepts (e.g., \textit{AI agent}) are added to the list, including their recognized definitions as research in the field progresses \cite{iso_22989_2022}.

\textbf{\challengeOpp{Standards Are Specifications-Driven}} The core DNA of standards is its technical specifications, which detail precise descriptions of the inputs and outputs of products being measured and the expected behaviors of processes. This approach parallels established engineering disciplines, where clear, non-ambiguous specifications enable the development of modular and robust systems 
\cite{weber1986specification,stoica2024specifications}. The quantity of specifications that make up a complete standard depends on the standard's complexity. The challenge for GenAI models is to learn these specifications and apply them dynamically or on an as-needed basis. Moreover, showing which specifications have been used by the GenAI model to generate content or check for compliance can fulfill regulatory demands for transparency and improve user trust 
\cite{ehsan2019automated,wei2022chain,liu2023increasing}. 

\textbf{\challengeOpp{Standards Have Limited Reference Data}} The clarity and level of information needed by domain experts in interpreting input and output requirements and learning from a few standard-compliant examples may not be equivalent to what GenAI models require. This limitation may pose challenges when automating compliance or conformity assessment using GenAI models. Some standards, especially those focusing on prescribed semantic and syntactic information, such as MLCommons' AILuminate Standard for assessing safety responses of LLMs \cite{mlcommons_ailuminate} or the NHS Standard for creating health-related content \cite{nhs_health_content_standard}, may have specifications where only a few conforming examples (typically 2-3) are provided. For cases like these, researchers often need to conduct their own data collection from external resources or perform synthetic data generation to increase the gold-standard reference data for compliance-based tasks, which may entail additional costs. These practices have been explored in previous works on using GenAI for compliance assessment concerning the HIPAA Privacy Rule and GDPR Documentation Standard \cite{colombo2024saullm,zoubi-etal-2024-privat5,fan-etal-2024-goldcoin}. 

\textbf{\challengeOpp{Standards Depend on Domain Knowledge Expertise}} Standards developed by experts in fields such as healthcare, science, and engineering require domain knowledge to interpret how their specifications and constraints can be applied to domain-specific processes or activities that require compliance. Baseline GenAI models are often trained with massive collections of web-scraped internet data combined with scaling techniques from which it can learn generic, jack-of-all-trades knowledge required for most tasks \cite{chung2024scaling,chowdhery2023palm}. LLMs, in particular, often demonstrate better performance than their baseline counterparts in domain-specific tasks when further finetuned or optimized with additional curated datasets that provide deeper domain knowledge. \cite{chen2023meditron}. This can be evidenced by previous works such as the Meditron-70B model, where they finetuned a Llama-70B with 40K clinical medical standards and guidelines from online healthcare websites and over 16M medical abstracts and papers from PubMed and PubCentral, which obtained higher performance across medical QA tasks than other closed and open-weight LLMs \cite{chen2023meditron}. This static finetuning of large models may be effective only for well-established standards that will not be updated for a substantial amount of time but may not be suited for standards, policies, or guidelines that are inherently dynamic and flexible. 

\textbf{\challengeOpp{Standards Require Strong Expert-Level Evaluation}} Perhaps one of the most important aspects of aligning GenAI with standards for compliance tasks is how we evaluate such alignment in terms of accuracy and practical usability. As discussed in Section~\ref{sec:paradigm_shift}, conformity assessments with standards can be a variable process depending on the level of compliance required for inputs and on how protocols work across different domains and sectors. Thus, the process of evaluation for standard compliance tasks should anchor to an agile, use-case basis that targets specific output requirements rather than adopting a one-size-fits-all approach \cite{khanzada2024conformity}. Ultimately, domain experts in standards should already be involved even at the conceptualization stage of GenAI-based workflows related to standard compliance to help identify plausible evaluation metrics to improve reliability and usability in real-world settings.

\subsection{Advantages and Benefits of Aligning GenAI with Standards}

\textbf{\positiveOpp{Standard Alignment Can Enhance Quality and Interoperability}} GenAI models are often controlled through various experimental means, such as different forms of structured prompting \cite{brown2020language,shin-etal-2020-autoprompt,wei2022chain}, finetuning \cite{weifinetuned2022,zhang2023adding,chowdhery2023palm}, and preference optimization \cite{ziegler2019fine,ouyang2022training,rafailov2024direct}, which have shown effectiveness across various tasks and domains. In line with this, standards can serve as a \textit{reference of control} for these models to generate and refine their outputs based on the standard's specifications, as done in recent works on education and language proficiency assessment \cite{imperial-etal-2024-standardize} and safe response generation using company policies \cite{guan2024deliberative}. Likewise, in relation to the emerging body of work with GenAI-based agents, aligning them with standards to produce an interconnected ecosystem can enable enhanced interoperability between inputs and outputs, thus improving efficiency, transparency, and the production of quality-controlled content.

\textbf{\positiveOpp{Standard Alignment Can Improve Oversight, Transparency, and Auditing}} In high-stakes domains, human oversight of any AI-based system or interface is crucial for transparency and auditing 
\cite{bowman2022measuring,kentonscalable}. As such, controlling how GenAI models produce outputs by updating the specifications of standards and being able to trace back deviations through these changes will be extremely valuable in areas such as bias mitigation \cite{gallegos2024bias}, fairness evaluations 
\cite{teo2024measuring}, and domain-specific tasks related to healthcare, finance, and legal decision-making 
\cite{bowman2022measuring,mesko2023imperative,mokander2023auditing}. Likewise, this form of oversight achieved by controlling with standards can also be scaled through \textit{superalignment}\footnote{First coined by OpenAI: \url{https://openai.com/index/introducing-superalignment/}} where a smaller GenAI teacher model specializing in a particular standard can regulate larger student models while exploiting its enhanced instruction-following capabilities \cite{burns2024weak,guo2024vision}. 

\textbf{\positiveOpp{Standard Alignment Can Strengthen User Trust}} Building user trust is considered one of the most elusive challenges in the design of AI-based systems, as it can dictate the lifeline of how these systems will be adopted and used \cite{riegelsberger2005mechanics,kizilcec2016much,schmidt2020transparency}. Bridging the same rules and regulations that domain experts follow, in the form of standards to control GenAI models, can potentially enhance process-based user trust. For example, a medical expert may feel much more confident in using a specialized open model like Meditron \cite{chen2023meditron} to complete or assist with their tasks than in using a black-box general-purpose model, simply from knowing that the former has undergone further training using massive collections of clinical guidelines and medical papers with which the expert is familiar. This level of transparency given to domain experts can be highly beneficial for earning and strengthening user trust, as it assures them of certainties in the performance expectations and design of standard-aligned GenAI models \cite{kizilcec2016much}.


\textbf{\positiveOpp{Standard Alignment Can Reduce Risk of Inaccuracies}}  
The results of the State of AI in 2024 survey conducted by McKinsey revealed that \textit{inaccuracy}---the tendency of GenAI models to produce factually incorrect and unexpected results---as one of the highest risk factors hindering adoption across major organizations and industries. Such risks can cause a domino effect, including losing user trust, potential physical or mental harm, and financial losses for both consumers and businesses if not properly mitigated and controlled \cite{mckinsey2024stateofai}. In line with this, the standard alignment of GenAI models can contribute to reducing inaccuracies by pairing it with architectural enhancements such as scaling and finetuning with massive collections of domain-specific data to improve domain knowledge, as evidenced in previous works on education \cite{imperial-tayyar-madabushi-2023-flesch,imperial-etal-2024-standardize}, legal \cite{zhu2024legilm,fan-etal-2024-goldcoin}, and medicine \cite{chen2023meditron}. Additionally, businesses can also use standard alignment as proof that their services conform to domain-specific regulations, thereby providing quality assurance to clients and consumers.

\section{Risks, Responsibilities, and Recommendations}
\label{sec:recommendations}

At this stage, we outline the potential risks and key responsibilities of each stakeholder group involved in the development of standards and advancement of GenAI and provide recommendations to align with the evolving practices in standard compliance.

\textbf{\recommendationOpp{For Government and Regulatory Bodies}} Regulations established by governing bodies can be considered one of the major drivers of standard development (see Section~\ref{sec:background}). However, two of the main concerns with regulations are the \textit{risk of rigid standardization} which may compromise innovation efforts and hinder or slow down the beneficial applications of AI \cite{aghion2023impact}, and the \textit{risk of regulatory gaps} due to vague stipulations and the unrealistic technical feasibility of legislations \cite{pouget2023}. To address these, we recommend periodic \textit{regulatory and policy adaptation} to revisit existing regulations and critically discuss how GenAI's evolving role impacts compliance practices with these regulations. Likewise, legislation aimed at developing a unified \textit{meta-standard} can also be enacted to assist professional regulatory bodies in effectively updating their professional guidelines for specific workforces that will experience major job augmentations with the use of GenAI (e.g., physicians now using GenAI to support clinical decision-making for patients in the UK), in order to uphold ethical principles and ensure human oversight \cite{Hashem2024}. Lastly, since the challenge of regulatory compliance through standards is closely tied to advanced research on GenAI, we propose ongoing research funding for academic and industry partners to promote scientific advancements in accountability, responsibility, and transparency of GenAI.

\textbf{\recommendationOpp{For SDOs, Industry, and Academic Expert Groups}} Rethinking the conventional standard development process (see Figure~\ref{fig:standard-hierarchy}) due to the shift in compliance practices with GenAI might be a major but necessary step for SDOs, industry associations, and academic expert groups. This is important to prevent the regulatory authority of technical standards from the \textit{risk of being uninformative} for users wanting clear guidelines for standard compliance amid the progress of GenAI. We recommend establishing a dedicated committee responsible for updating and initiating maintenance revisions of existing published standards. These domain-specific committees can conduct their own studies and collaborate with other stakeholders to \textit{closely monitor current practices, available tools, and limitations} in using GenAI for standard compliance \cite{manheim2024necessity}.  Similarly, to boost advancements in GenAI research with standard compliance, we recommend open-sourcing machine-readable format of standards along with producing gold-standard compliant data for the research community.

\textbf{\recommendationOpp{For GenAI Researchers and Model Developers}} When an AI model is deployed in critical high-risk domains such as healthcare, legal, or engineering safety, it is the responsibility of researchers and developers to propose explainability techniques to interpret model outcomes. This aspect of GenAI research is important to avoid the \textit{risk of eroding trust} among domain users who will use this technology. Thus, we recommend using standard compliance as an impactful case study of \textit{explaining} the black-box nature of GenAI models. Since standards can be considered as co-regulation tools, developing novel approaches such as automatic audit trail generation for decisions made by standard-aligned GenAI and providing human-readable explanations is vital for its wider interdisciplinary adoption \cite{mokander2023auditing, song2019auditing}. In addition, we also recommend that researchers collaborate with domain experts and explore using standards and regulatory documents as \textit{references for control} for content generation tasks. This research direction will contribute to realistic applications of the capabilities of GenAI where the task of standard compliance with documents can be added to LLM benchmark evaluation suites such as HELM \cite{liang2023holistic}, ChatBot Arena \cite{chiang2023chatbot}, and BIG-Bench \cite{srivastava2023beyond}. 

\textbf{\recommendationOpp{For Regulated Entities, Practitioners, and Users}} The final stakeholder group that will experience the greatest impact from GenAI are the regulated entities, practitioners, and users. Our most important recommendation is to practice and uphold the highest form of responsibility and accountability in using GenAI to comply with regulatory and operational requirements \cite{coeckelbergh2020artificial}. Professionals in regulated fields do not need to understand the full technical inner workings of GenAI. However, to reduce the \textit{risk of misuse and over-reliance}, they should remain knowledgeable about the limitations of any GenAI model, including its tendency to produce inaccurate responses and exhibit limited domain expertise. Thus, we strongly recommend establishing a \textit{solid, domain-specific foundation of AI literacy}, which is expected of an ideal professional who knows how to work with and maximize the potential of these intelligent tools. Finally, we recommend close collaboration with all stakeholders in the standard development process while providing feedback to enhance the overall usability and experience of standard compliance with GenAI models.

\section{Alternative Views}
\label{sec:alternative}

While we emphasized and supported our position in the previous sections, we present two main alternative views that we consider equally valid and necessary to ensure a healthy discussion in this emerging new research direction. 

\textbf{Aligning GenAI for Compliance May Be Superficial} GenAI can produce non-factual, hallucinated responses if it lacks sufficient domain knowledge for various tasks. Deploying GenAI-based assistants, particularly those classified below the Adaptive level in compliance capabilities in \textsc{C3F}, is not realistic since AI often struggles with \textit{very specific edge cases and context-dependent ambiguities} that only experienced domain experts can realistically resolve. In response to this, we believe GenAI is \textit{not} intended to take a significant portion of work from domain experts, but rather, to act as \textit{first-line assistants} in managing automatic, repetitive preliminary tasks, thereby allowing domain experts to focus other parts of their work. We emphasize that maintaining \textit{human oversight} is crucial for any AI-based workflow, as indicated in the critical levels assessment of standards in \textsc{C3F}. Using GenAI as first-line assistants is also viable for standard-compliance tasks requiring generating content that conforms to specific writing conventions (e.g., ASD-STE Simplified Technical English \cite{asd-ste100-issue9,imperial-tayyar-madabushi-2024-specialex}) and can be easily verified with rule-based software. 

\textbf{Aligning GenAI for Compliance Creates False Sense of Security} Optimizing GenAI to follow rigid standards might result in users and practitioners being too \textit{overreliant} on these systems. Even if users are domain experts themselves, they can subjected to a \textit{false notion of security} from an over-confident model, which might lead to reduced human vigilance and possibly cause major to catastrophic errors from non-compliance. In response to this, we acknowledge that \textit{trust is a multifaceted concept}, particularly for AI systems \cite{papenmeier2022s}. As discussed in Section~\ref{sec:recommendations}, our position on aligning GenAI with standards entails the need for strong collaborative efforts across all stakeholders in designing \textit{user-centric approaches}. This recommendation is crucial as it enables regulated entities, practitioners, and users to actively engage in the \textit{process of standard alignment} for GenAI models, where they will understand the importance of human oversight and accountability in using these systems and the potential risks of overreliance.

\section{Outlook}
\label{sec:conclusion}

GenAI is transforming how we work and perform day-to-day tasks and will continue to do so in the near future. In this paper, we view such transformation as a \textit{paradigm shift} and discuss how supporting this change can be considered \textit{a fundamental step toward building trustworthy, controllable, and responsible GenAI systems} through standard alignment. 

Based on our position, we make \textbf{two calls to action} across all stakeholder groups involved in the development of standards and GenAI:

\begin{enumerate}
    \item We encourage all stakeholders to collectively recognize and adapt to the paradigm shift introduced by GenAI, which can serve as a powerful tool for assisting with compliance to regulations and standards, provided it is used responsibly and ethically.
    \item We advocate for a global, unified effort among stakeholders to fully realize the potential of this research direction and ensure that GenAI remains a force for good in regulated domains.
\end{enumerate}

We look forward to welcoming new research ideas, global partnerships, initiatives, and collaborations, whether they complement or contrast with the position we introduced, in line with the promising direction of aligning GenAI for regulatory and operational compliance through standards.

\begin{ack}
We are grateful to a number of people who contributed to the discussions on the position and provided constructive feedback on earlier versions of this paper: Orlando Chiarello (ASD-STE STEMG), Howard Benn (ETSI), Julian Padget (University of Bath, IEEE P7003 WG), James Davenport (University of Bath, BSI), Tory Frame, Matthew Hewitt, Marina De Vos, George Fletcher, and Jinha Yoon. JMI is supported by the National University Philippines and the UKRI Centre for Doctoral Training in Accountable, Responsible, and Transparent AI [EP/S023437/1] of the University of Bath.
\end{ack}


\bibliographystyle{abbrvnat}
\bibliography{references.bib}

\clearpage
\appendix


\section{Hierarchy of the Standard Development Process}
\label{app:hierarchy}

\begin{figure}[!htbp]
    \centering
    \includegraphics[width=0.50\linewidth]{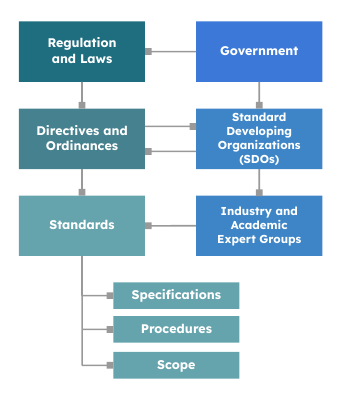}
    \caption{A supporting visualization of the general process of developing standards. Standards can be created either by government and regulatory bodies as a product of legislation or through industry associations and academic expert groups to ensure interoperability and quality of systems and processes.}
    \label{fig:standard-hierarchy}
\end{figure}

\section{Technical Enhancements for Standard Alignment}
\label{app:technical}

We provide supplementary information on promising novel technical advancements that can be explored (but are not limited to) to further improve how GenAI models can align with standards and cater to their inherent complexities as discussed in Section~\ref{sec:challenges}. Some of these approaches may have been preliminarily explored by recent works for selected domains with available machine-readable standards data. Collaborations between AI and interdisciplinary areas can foster further advancements and open more novel approaches toward standard alignment.

\textbf{Constraint and Knowledge Representations}
Specifications from standards can be represented as a form of a structured set of constraints either as input or as part of an embedded training regime for aligning GenAI models. The simplest example of this is through in-context learning (ICL), where constraints are framed as prompts in an instruction-like manner with specific informative examples provided to show the target output required from the model
\cite{brown2020language,mishra-etal-2022-reframing}. This has been done on works such as the production of high-quality standard-conforming educational content with CEFR and Bloom's Taxonomy \cite{imperial-etal-2024-standardize,malik-etal-2024-tarzan,elkins2024teachers} and rewriting complex texts to conform to government-mandated plain languages guidelines \cite{da2022redactor,joseph-etal-2024-factpico}. The advantage of ICL is its simplicity, which can easily be explored by non-technical domain users with any arbitrary GenAI-based chat interface (e.g., ChatGPT) and can be attributed to how the paradigm shift started (see Section~\ref{sec:paradigm_shift}). More advanced levels of representation focus on transforming standards into knowledge graphs and ontologies. An example of this is the work by \citet{hernandez2024open}, where they transformed the text content of the EU AI Act into a high-level knowledge graph to show links between defined terms and their associated requirements from the Act's statements for compliance checking.

\textbf{Post-Training Improvements}
Post-training has become one of the major foci in ML research since the release of preference optimization techniques like RLHF \cite{ouyang2022training} and DPO \cite{rafailov2024direct} combined with supervised finetuning (SFT) to improve the response alignment of GenAI models with respect to task requirements. All Baseline models in Table~\ref{tab:model_classification} have undergone post-training through SFT. For standard alignment, post-processing can potentially be emulated by aggregating prompt and response pairs that conform to the specifications of a standard and then finetuning a pretrained model with this data. An example reference for this is OpenAI's Deliberative Alignment method \cite{guan2024deliberative} where an LLM is trained with chain-of-thought (CoT) style prompts \cite{wei2022chain} can identify whether which safety policy specification is applicable when identifying whether to respond to a prompt or not. However, for researchers who want to explore this approach, one caveat is that it will require at least 1,000 instances of very high-quality, expert-level pairwise data to achieve relatively decent performance \cite{zhou2023less}. Nonetheless, combinations of post-training techniques (e.g., SFT + DPO with standard-aligned preference pairs + CoT prompting) are viable approaches for improving a GenAI model's standard compliance capabilities.

\textbf{Synthetic Data Generation}
Synthetic data generation using modern GenAI models that have undergone processes such as larger scaling, instruction-tuning, and preference optimization often outperform other data augmentation techniques for downstream tasks \cite{ye-etal-2022-zerogen,li-etal-2023-synthetic}. In standard alignment of GenAI models, researchers can explore several options related to synthetic data generation to improve compliance capabilities. First, in parallel with post-training enhancements from above, using synthetic data in the form of standard compliant and non-compliant examples can be a practical choice to optimize a model's generation qualities. This approach has been applied by \citet{fan-etal-2024-goldcoin}, where they generated synthetic case scenarios for GDPR and HIPAA Privacy Rules to finetune smaller LLMs for compliance detection. Moreover, recent works have documented higher performance for models that have been finetuned with a combination of high-quality expert data and machine-generated data in alignment tasks which makes the process of compiling standard-aligned feedback data relatively easier \cite{miranda2024hybrid,ivison2024unpacking,lee2024rlaif}.

\textbf{Retrieval and Tool Augmentation}
Augmenting GenAI models, particularly LLMs, with external tools to enhance their problem-solving capabilities has gained increasing research attention in recent years. The use of tools such as calculators, search engines, and API function calls has been shown to improve the zero-shot performance of LLMs across question-answering downstream tasks requiring up-to-date information \cite{hao2023toolkengpt,schick2023toolformer}. In the case of standard alignment, syntactic content-based standards such as the CEFR and CCS standards (see Table~\ref{tab:standards_classification} for reference) requiring specific characteristics of texts such as sentence lengths to measure complexity can greatly benefit from a GenAI model that knows how to call a calculator tool to approximate how long sentences should be generated. Moreover, a search engine tool can also help GenAI models access updated versions of standard specifications from its original web sources as prior version checking. On the other hand, another approach that can improve GenAI models' domain knowledge is to encapsulate it in a retrieval-augmented generation (RAG) ecosystem where auxiliary retrievers that have access to external knowledge bases that can be added to as context to prompts \cite{lewis2020retrieval,ram-etal-2023-context}. 

\textbf{Reasoning Capabilities}
Whether GenAI models, such as LLMs, can reason or not is a highly debated topic in current ML research. Reasoning is an inherent ability that plays a crucial role in how humans solve problems through critical thinking \cite{huang-chang-2023-towards}. Previous research often claims that such capability can be triggered in different ways, such as providing intermediary reasoning steps to prompts \cite{wei2022chain} or using arbitrary models to select and infer reasoning steps from context information \cite{creswell2022faithful}. Assuming GenAI models can actually reason, in standard alignment, such skill may play an important role in deciding which specifications of a standard are required to be followed and which ones can be disregarded safely, given the additional context information of a task. Preliminary work in this direction includes OpenAI's Deliberative Alignment method \cite{guan2024deliberative} where an LLM is optimized to reason over which safety policy specifications should be followed and which can be ignored using their o-series models. As such, GenAI models rated Advanced and Adaptive in \textsc{C3F} should document convincing reasoning capabilities across applicable tasks.

\section{Full Assessment Results of GenAI Models and Standards with \textsc{C3F}}
\label{app:framework_tables}

As discussed in Section~\ref{sec:framework}, we provide the full list of the 15 selected foundational and domain-finetuned GenAI models assessed based on their compliance capabilities and the 34 selected standards across multiple disciplines for their criticality levels.

\onecolumn

\definecolor{lightgray}{RGB}{217, 215, 213}  
\definecolor{lightblue}{RGB}{230, 235, 245}     
\definecolor{lightpurple}{RGB}{240, 230, 245}   

\newcommand{\subscription}{Subscription}
\newcommand{\openweight}{Open Weight}
\newcommand{\opencode}{Open Code}

\renewcommand{\arraystretch}{1.5}
\begin{table}[!t]
\small
\centering
\begin{tabular}{@{}
  >{\raggedright\arraybackslash}p{3.3cm}
  >{\raggedright\arraybackslash}p{2.5cm}
  >{\raggedright\arraybackslash}p{2.3cm}
  >{\raggedright\arraybackslash}p{1.8cm}
  >{\centering\arraybackslash}p{2cm}
  @{}}
\toprule
\textbf{\textsc{Model}} & 
\textbf{\textsc{Organization}} & 
\textbf{\textsc{Domain}} & 
\textbf{\textsc{Openness}} & 
\textbf{\textsc{Compliance}} \\ \midrule
\textsc{o-series}                & OpenAI                & General                 & \subscription              & \advanced    \\
\textsc{GPT-4}             & OpenAI                & General                 & \subscription              & \specialized    \\
\textsc{DeepSeek-R1}                & DeepSeek-AI                & General                 & \openweight              & \baseline    \\
\textsc{Claude Opus}       & Anthropic             & General                 & \subscription              & \baseline    \\
\textsc{Gemini 2.0}        & Google                & General                 & \subscription              & \baseline    \\
\textsc{Llama 3.1 405B}    & Meta                  & General                 & \openweight               & \baseline    \\
\textsc{Mistral Large}     & Mistral               & General                 & \subscription & \baseline    \\
\textsc{Command-R 105B}    & Cohere                & General                 & \openweight & \baseline    \\
\textsc{Midjourney 6.1}    & Midjourney                & General                 & \subscription & \baseline    \\
\textsc{DALL-E}    & OpenAI                & General                 & \subscription & \baseline    \\\midrule
\textsc{Meditron 70B}      & \citet{chen2023meditron}     & Healthcare              & \opencode                 & \specialized \\
\textsc{GoldCoin-Llama}    & \citet{fan-etal-2024-goldcoin}      & Healthcare, Legal       & \opencode                 & \specialized \\
\textsc{LegiLM}            & \citet{zhu2024legilm}      & Legal                   & \opencode                 & \specialized \\
\textsc{ChemCrow}          & \citet{m2024augmenting}     & Chemistry               & \opencode                 & \specialized \\
\textsc{Standardize-Llama} & \citet{imperial-etal-2024-standardize} & Education               & \opencode                 & \specialized \\ 

\bottomrule
\end{tabular}
\vspace{10pt}
\caption{We used the \textbf{compliance capabilities} component of \textsc{C3F} in Figure~\ref{fig:framework} to assess the latest versions of 15 foundational and specialized GenAI models both for text and image. We also include information on their respective domains (in the case of Specialized models) and accessibility (Open Weight, Open Code, or Subscription). The first section of the table includes industry-released GenAI models (mostly Baseline except for o-series models by OpenAI) while the second section is more focused on Specialized models commonly led by academic and research groups targeting specific domains and sectors. Models and their compliance capabilities in this table serve as a non-exhaustive example and can be extended or re-assessed as supporting literature are released.}
\label{tab:model_classification}
\end{table}


\small
\renewcommand{\arraystretch}{1.5}

\begin{longtable}{@{}
  >{\raggedright\arraybackslash}p{4cm}
  >{\raggedright\arraybackslash}p{2cm}
  >{\raggedright\arraybackslash}p{2.8cm}
  >{\raggedright\arraybackslash}p{1.5cm}
  >{\centering\arraybackslash}p{1.5cm}
  @{\hspace{10pt}}@{}}

\toprule
\textbf{\textsc{Standard}} &
  \textbf{\textsc{Domain}} &
  \textbf{\textsc{Organization}} &
  \textbf{\textsc{Openness}} &
  \textbf{\textsc{Criticality}} \\ \midrule
\endfirsthead

\toprule
\textbf{\textsc{Standard}} &
  \textbf{\textsc{Domain}} &
  \textbf{\textsc{Organization}} &
  \textbf{\textsc{Openness}} &
  \textbf{\textsc{Criticality}} \\ \midrule
\endhead

\bottomrule
\multicolumn{5}{r}{\textit{Continued on next page}} \\
\endfoot
\bottomrule
\caption{We used the \textbf{criticality levels} component of \textsc{C3F} in Figure~\ref{fig:framework} to classify a wide variety of standards across domains and sectors based on their sensitivity which translates into the margin of permissible errors a hypothetical GenAI model can potentially commit when assisting with standard compliance tasks. We include supporting information including the organization or initiative which led to the development of the standard as well as its accessibility (Public or Subscription). For standards classified from the domains of healthcare and engineering, we obtained direct recommended assessments from expert practitioners and professionals with respect to the criticality classification criteria in \textsc{C3F}.}\label{tab:standards_classification} 
\endlastfoot

AILuminate Assessing Safety Standard &
  Software and Technology &
  MLCommons &
  Public &
  \minimal \\
Simplified Molecular Input Line Entry System (SMILES) &
  Chemistry &
  David Weininger &
  Public &
  \minimal \\
Associated Press Stylebook &
  Media and Communications &
  Associated Press &
  Subscription &
  \minimal \\
Plain Language Standards &
  Media and Communications &
  (region-specific) &
  Public &
  \minimal \\
Text Encoding Initiative &
  Software and Technology &
  TEI Consortium &
  Public &
  \minimal \\
ISO/IEC Systems and Software Engineering Documentation &
  Software and Technology &
  International Electrotechnical Commission &
  Subscription &
  \minimal \\
FAIR Data Principles Documentation &
  Software and Technology &
  GO FAIR Initiative &
  Public &
  \minimal \\
OpenAPI Specification (OAS) &
  Software and Technology &
  SmartBear Software &
  Public &
  \minimal \\
Section 508 Compliance Documentation &
  Software and Technology &
  US Government &
  Public &
  \minimal \\
Python Enhancement Proposals &
  Software and Technology &
  Python &
  Public &
  \minimal \\
Web Standards &
  Software and Technology &
  W3C &
  Public &
  \minimal \\
ASD-STE100 Simplified Technical English (STE) &
  Engineering &
  Aerospace, Security and Defence Industries Association of Europe &
  Public &
  \moderate \\
Common European Framework of Reference for Languages (CEFR) &
  Education &
  Council of Europe &
  Public &
  \moderate \\
Common Core Standards (CCS) &
  Education &
  National Governors Association, Council of Chief State School Officers &
  Public &
  \moderate \\
Preferred Reporting Items for Systematic Reviews and Meta-Analyses (PRISMA) &
  Healthcare &
  PRISMA Initiative &
  Public &
  \moderate \\
Standards for Quality Improvement Reporting Excellence (SQUIRE) &
  Healthcare &
  SQUIRE Initiative &
  Public &
  \moderate \\
IFRS Accounting Standards &
  Finance &
  International Financial Reports Standards &
  Public &
  \moderate \\
Systematized Nomenclature of Medicine Clinical Terms (SNOMED CT) &
  Healthcare &
  National Health Service &
  Public &
  \high \\
NHS Health Content Standards &
  Healthcare, Government &
  National Health Service &
  Public &
  \high \\
HIPAA Privacy Rule &
  Healthcare, Government &
  US Government &
  Public &
  \high \\
GDPR Documentation Standard &
  Legal, Government &
  European Union &
  Public &
  \high \\
International Classification of Diseases (ICD) Standard &
  Healthcare, Government &
  Centers for Disease Control and Prevention &
  Public &
  \high \\
Strengthening the Reporting of Observational Studies in Epidemiology (STROBE) Guidelines &
  Healthcare &
  STROBE Initiative &
  Public &
  \high \\
Case Report Guidelines (CARE) &
  Healthcare &
  CARE Initiative &
  Public &
  \high \\
Standard Protocol Items: Recommendations for Interventional Trials (SPIRIT) &
  Healthcare &
  SPIRIT Initiative &
  Public &
  \high \\
Digital Imaging and Communications in Medicine (DICOM) &
  Software and Technology, Healthcare &
  National Electrical Manufacturers Association &
  Public &
  \high \\
USDA Food Safety Documentation &
  Healthcare, Government &
  US Government &
  Public &
  \high \\
AGREE Reporting Checklist &
  Healthcare &
  International Appraisal of Guidelines, Research and Evaluation (AGREE) &
  Public &
  \high \\
NICE Process and Methods &
  Healthcare &
  National Institute for Health and Care Excellence &
  Public &
  \high \\
Operational Design Domains &
  Engineering &
  (company-specific) &
  Public &
  \high \\
BBC Content Standards &
  Media and Communications &
  Office of Communications &
  Public &
  \high \\
Consolidated Standards of Reporting Trials (CONSORT) &
  Healthcare &
  CONSORT Group &
  Public &
  \high \\
IAEA Safety Standards &
  CBRN &
  International Atomic Energy Agency &
  Public &
  \extreme \\
Responding To A CBRN Event: Joint Operating Principles for the Emergency Services &
  CBRN &
  Joint Emergency Services Interoperability Programme &
  Public &
  \extreme 
\\      

\end{longtable}
\clearpage 
\twocolumn

\end{document}